\documentclass{amsart}
\usepackage{amssymb}
\usepackage{verbatim}
\usepackage{mathrsfs}
\usepackage{dsfont}
\usepackage{braket}

\theoremstyle{plain}
\newtheorem{thm}{Theorem}[section] 
\theoremstyle{definition}
\newtheorem{defn}[thm]{Definition} 
\newtheorem{rem}[thm]{Remark}
\newtheorem{con}[thm]{Conjecture}
\newtheorem{prop}[thm]{Proposition}
\newtheorem{lem}[thm]{Lemma}
\newtheorem{exmp}[thm]{Example}

\newtheorem{cor}[thm]{Corollary}

\newcommand{\id}{\mathrm{id}}

\newcommand{\cH}{\mathcal{H}}

\newcommand{\cA}{\mathcal{A}}
\newcommand{\cB}{\mathcal{B}}
\newcommand{\cP}{\mathcal{P}}
\newcommand{\cT}{\mathcal{T}}

\newcommand{\Tr}{\mathrm{Tr}}
\newcommand{\Cn}{{\setbox0=\hbox{
$\displaystyle\rm C$}\hbox{\hbox
to0pt{\kern0.6\wd0\vrule height0.9\ht0\hss}\box0}}} 

\newcommand{\jed}{{\mathbb{I}}}
\begin{document}
\title{On PPT square conjecture.} 

\author{Wladyslaw Adam Majewski}

\email{fizwam@gmail.com}

\address{DSI-NRF CoE in Math. and Stat. Sci,\\ Focus Area for Pure and Applied Analytics,\\ Internal Box 209, School of Comp., Stat., $\&$ Math. Sci.\\NWU, PVT. BAG X6001, 2520 Potchefstroom\\ South Africa}


\date{\today}

\subjclass[2010]{}

\keywords{positive maps, tensor cones, PPT conjecture, selection rule}
\begin{abstract}
An analysis of the PPT square conjecture
based on the characteristics of positive mappings given in \cite{Maj} will be carried out.
First, a general strategy for analyzing the PPT conjecture will be presented. Then, a detailed analysis of the certain tensor cones and their corresponding classes of mappings will be given. Special emphasis will be put on key mappings used in quantum information.
\end{abstract}

\maketitle

\newpage
\section{Introduction}
The notion of positive partial transpose maps, PPT maps for short, is playing an important role in the quantum information theory. We remind the reader that linear maps $T: B(\cH) \to B(\cH)$  that are both completely positive and completely copositive are called PPT maps. 
The important point to note here is an observation  that  various classes of  positive maps are used to describe quantum and useful maps in bipartite systems. In particular, quantum maps which are useless for preserving quantum entanglement are called entanglement breaking maps. $T$ is an entanglement breaking map iff it can be written as
\begin{equation}
\label{super}
T(a) = \sum_k s_k(a) P_k,
\end{equation}
where $s_k(a) = (a w_k, w_k)$, $a \in B(\cH)$, $P_k = |v_k><v_k|$ for some sets of vectors $w_k, v_k$ in $\cH$; see \cite{HSR}, \cite{KMP}.

Recently, Christandl, \cite{R}, raised the following conjecture, see also \cite{B}, \cite{BCHW}, \cite{KS}:
\begin{con} (PPT square conjecture)
\label{1}
The composition $T_1 \circ T_2$ of a pair of
$T_1$ and $T_2$ PPT maps on $B(\cH)$ is always  entanglement breaking.
\end{con}

It was argued that PPT square conjecture \ref{1} is equivalent to
\begin{con} \cite{ChMW}
\label{2}
If $T_1$ is positive and $T_2$ is PPT map then $T_1 \circ T_2$ is decomposable.
\end{con}

Based on the characteristics of positive mappings, which was given in \cite{Maj}, the above hypotheses will be analyzed in detail. The paper is organized as follows.
In Section 2 we give  definitions of the concepts just introduced. Furthermore, we will define and discuss the associated classes of positive maps.  Section 3 is devoted  to a study of PPT square conjecture.  In particular, we will present the general approach to the problem. Keeping in mind that entanglement breaking maps are so important in Quantum Information, in Section 4 we will give a detailed analysis of the relevant tensor cones  and associated classes  of maps. This will be done  for finite dimensional cases. 
 
\section{Preliminaries}

To facilitate access to the considered maps,  here, we will make the following assumptions:  let $\{e_i\}_1^{d < \infty}$ be a basis in $\cH$, where $d = \dim \cH$. The conjugation $J: \cH \to \cH$ is defined as
\begin{equation}
J f = J \left(\sum_i (f,e_i) e_i\right) = \sum_i \overline{(f,e_i)} e_i,
\end{equation}
where $f\in \cH$. In particular, a vector  $f\in \cH$ is called real if $Jf =f$. The transposition $t(\cdot)$ of $a \in B(\cH)$ with respect to the basis $\{e_i\}$ is given by
\begin{equation}
t(a) = J a^* J.
\end{equation}

Our next observation is that any $V \in B(\cH)$ can be written as
\begin{equation}
Vf = V\left(\sum_k (f,e_k)e_k\right) = \sum_k (f,e_k) Ve_k = \sum_k
\left(Ve_k \otimes \overline{e_k}\right) f,
\end{equation}
where  $f \in \cH$, and $v \otimes \overline{w} 
\ \cdot z = (z,w) v$; for details see \cite{Sch}.

In the very special case, if $Ran V = 1$ one has $Ve_i = \lambda_i x$, where $x$ is a certain vector in $\cH$ and $\lambda_i \in \Cn$. Moreover, in this case,
\begin{equation}
V = \sum_i \lambda_i x \otimes \overline{e_i} = x \otimes \overline{\sum_i \overline{\lambda_i} e_i} = x \otimes \overline{y}.
\end{equation}
  
Thus,  (\ref{super}) can be rewritten as
\begin{equation}
\label{super2}
T(a) = \sum_k v_k \otimes \overline{w_k} \cdot a \cdot w_k \otimes \overline{v_k} \equiv \sum_k V_k \cdot a \cdot V^*_k,
\end{equation}
and, obviously, $Ran v_k \otimes \overline{w_k} \equiv Ran V_k = 1$.

However, as a quantum system demands infinite dimensional Hilbert space, from now on we make the assumption that Hilbert spaces $\cH$ is, in general, infinite dimensional.
We need to distinguish the following families of maps $T: B(\cH) \to B(\cH)$;  positive, decomposable, completely positive, PPT, and superpositive. To present the concise and rigorous description we will use the selection rule which gives the required characterization in terms of tensor cones, see \cite{Maj} for all details.

\begin{itemize}
\item positive maps; they are determined by the tensor cone 
$conv\left(B(\cH)_+ \otimes \mathcal{F}_T(\cH)_+\right) = C_p$. Here and subsequently we are using the identification $\mathcal{F}_T(\cH) = B(\cH)_*$. Here,  $\mathcal{F}_T(\cH)$ stands for the trace class operators which is the predual space $B(\cH)_*$ with respect to $B(\cH)$.
\item decomposable maps; they are determined by the tensor cone 
$$conv\{ \left(B(\cH) \otimes \mathcal{F}_T(\cH) \right)_+ \cap \left(\id\otimes t \right) \left(B(\cH) \otimes \mathcal{F}_T(\cH) \right)_+\} \equiv C_d.$$
\item completely positive maps; they are determined by the tensor cone 

$\left(B(\cH) \otimes \mathcal{F}_T(\cH) \right)_+ \equiv C_{cp}$.
\item PPT maps; they are determined by the tensor cone 
$$conv\{\left(B(\cH) \otimes \mathcal{F}_T(\cH) \right)_+ \cup \left(\id\otimes t\right) \left(B(\cH) \otimes \mathcal{F}_T(\cH) \right)_+\} \equiv C_{PPT.}$$
\item super positive maps; they are determined by the largest tensor cone - the injective tensor cone $C_i$.
Equivalently, $T(a) = \sum_i V^*_i a V_i$ is superpositive iff there exists a Krauss decomposition such that for all $V_i$ one has the $Ran V_i = 1$, cf St{\o}mer's book \cite{Ster}.
\end{itemize}
 
We remind the reader that the selection rule is designed to define certain classes of positive maps with the prescribed properties, see  \cite{Maj}. It reads as follows: the class of positive maps  $P_{\alpha}$ is defined as
\begin{equation}
\label{SR}
P_{\alpha} = \{ T \in L(B(\cH), B(\cH)); \tilde{T}(\sum a_i \otimes b_i) \equiv \sum  Tr T(a_i) t(b_i) \geq 0 \ \rm{for \ any} \sum a_i \otimes b_i \in C_{\alpha} \}
\end{equation}
where $C_{\alpha}$ is the specified tensor cone. In other words, properties of a positive map $T$ are specified by the associated functional $\tilde{T}$, for all details see \cite{Maj}.

\section{PPT maps and PPT square conjecture}

To get some intuition we begin this section with an example.
\begin{exmp}
\label{3.1}
Let $V_1= v \otimes \overline{e_1} + Jv \otimes \overline{e_2}$,  $V_2= Jv \otimes \overline{e_1} + v \otimes \overline{e_2}$ where $v$ is neither a real  nor an imaginary vector in $\cH$, $\{e_i\}_{i=1}^2$ a basis in $\cH$. We remind that $v$ is a real vector if $v = Jv$ and an imaginary vector if $Jv = - v$.
Define
\begin{equation}
T(a) = \sum_{i=1}^2 V_i a V^*_i, \quad a \in B(\cH).
\end{equation}
Then
\begin{equation}
T(a) = (ae_1,e_1) v \otimes \overline{v} + (ae_2,e_1) v \otimes \overline{Jv} + (ae_1,e_2) Jv \otimes \overline{v}
+ (ae_2,e_2)Jv \otimes \overline{Jv}
\end{equation}
$$+ (ae_1,e_1) Jv \otimes \overline{Jv} + (ae_2,e_1) Jv \otimes \overline{v} + (ae_1,e_2) v \otimes \overline{Jv}
+ (ae_2,e_2)v \otimes \overline{v}, $$
and 
\begin{equation}
t \circ T(a) = T(a).
\end{equation}

Further, we note
\begin{equation}
T\circ T (a) = \sum_{ij} V_iV_j a V_i^* V_j^*,
\end{equation}
with
\begin{equation}
V_1^2 = \left(v\otimes \overline{e_1} + Jv \otimes \overline{e_2}\right)\left(v \otimes \overline{e_1} + Jv \otimes \overline{e_2}\right)
\end{equation} 
$$= (v,e_1) v\otimes \overline{e_1} + (e_1,v) v \otimes \overline{e_2} + (v,e_2) Jv \otimes \overline{e_1} + (e_2,v) Jv \otimes \overline{e_2} , \quad \rm{etc.}$$

It is clear that $Ran V_i = 2$ while $Ran V_i^2$ is also equal to $2$ if $v$ is not orthogonal to $\{e_1, e_2 \}$,  ($v \in \cH$ is not a real vector). Obviously, $T$ is a PPT map. In general, $T$ is not unital. However, there do exist such vectors $v$ (in the plane $<e_1,e_2>$ determined by vectors $e_1,e_2$) that $T(\bf{1}) = \bf{1}$.
Finally we want to see that $T$ is superpositive. 
The important point to note here is the fact that the Krauss decomposition, cf (\ref{super}), (\ref{super2}), is not unique. Therefore, to show that $T$ is superpositive we apply the characterization of these maps given in terms of the injective cone, see the previous section. To this end 
it would be enough to observe that, in 2-dimensional case, the tensor cone $conv\{\left(B(\cH) \otimes \mathcal{F}_T(\cH) \right)_+ \cup \left(\id\otimes t\right) \left(B(\cH) \otimes \mathcal{F}_T(\cH) \right)_+\}$
is equal to the injective cone $C_i$.  We will see that
 this claim  is dual to the equivalence, in two dimensional case, of the tensor cone $C_d$ determining decomposable maps with the tensor cone determining positive maps $C_p$.
Thus to proceed with the examination of superpositivity,  we should introduce the notion of duality and develop its description. Then, the superpositivity of $T$ will follow.
\end{exmp}

To define the dual tensor cone, firstly,  we will define the notion of duality, cf \cite{DF}. To this end we begin with an observation that the pair  $<B(\cH), \mathcal{F}_T(\cH)>$ of vector spaces $B(\cH)$ and $\mathcal{F}_T(\cH)$ is separating dual, i.e. 
\begin{equation}
a \times \varrho \mapsto <a, \varrho> \equiv \Tr a \varrho \in \Cn; \  a \in B(\cH), \  \varrho \in \mathcal{F}_T(\cH),
\end{equation}
and for any $a\neq0$ ($\sigma \neq0$) there is $\varrho \neq0$ ($b \neq0$) with $\Tr a \varrho \neq 0$ ($\Tr b \sigma \neq 0$ respectively), where $a,b \in B(\cH)$ and $\varrho, \sigma \in \mathcal{F}_T(\cH)$.

The separating dual pairing is defined as, cf Section 15 in \cite{DF},
$$ <\sum_n a_n \otimes \varrho_n, \sum_m b_m \otimes \sigma_m> \ \to \Cn$$
\begin{equation}
<\sum_na_n \otimes \varrho_n, \sum_m b_m \otimes \sigma_m> = \sum_n \sum_m \Tr a_n\sigma_m \cdot \Tr\varrho_n b_m.
\end{equation}

Now, we are in position to define 
\begin{defn}
The dual cone $C^d_{\alpha}$  is defined as
\begin{equation}
C^d_{\alpha} = \{ \sum_n a_n \otimes \varrho_n; \sum_m \sum_n \Tr a_n \sigma_m \cdot \Tr b_m \varrho_n \geq 0 \ \rm{for \ all} \ \sum_m b_m \otimes \sigma_m \in C_{\alpha} \}.
\end{equation}
\end{defn} 

In particular, it is an easy observation that
\begin{equation}
C^d_p = \{  \sum_n a_n \otimes \varrho_n; \sum_m \sum_n \Tr a_n \sigma_m \cdot \Tr b_m \varrho_n \geq 0, 
\ {for \ all} \ b_m \geq0, \sigma_m \geq 0 \} = C_i.
\end{equation} 

In addition, we note:
\begin{rem} 
\begin{enumerate}
\item  As tensor cones are defined  for  the projective tensor product, the above tensor products $\otimes$ denote  the projective tensor product $\otimes_{\pi}$.
\item Now it is an easy exercise to prove the final claim of the above example.
\item Finally, the above notion of duality suggests that two conjectures (\ref{1}) and (\ref{2}) are dual each other.
 We return to this issue at the end of this section.
\end{enumerate}
\end{rem}

We now turn to an  explanation of  what is going on with the composition of a positive map with a PPT map. This will be done using the structure of positive maps, and in particular, the selection rule; for  details see \cite{Maj}.

Our first observation is that Conjecture \ref{2} can be rephrased as follows: for any positive map $T_1$ the functional  $\tilde{T_0}$ associated with the restriction $T_0 \equiv T_1|_{T_2 C_d}$, where $T_2$ is a PPT map, should be positive on the tensor cone $\left(B(\cH) \otimes \mathcal{F}_T(\cH)\right)_+ \cap \left(id \otimes t\right)\left(B(\cH) \otimes \mathcal{F}_T(\cH)\right)_+ \equiv C_d$.

The associated functional $\tilde{T_0}$ to the map $T_0$ has the form, cf \cite{Maj}
\begin{equation}
\tilde{T_0}(\sum_i a_i \otimes b_i) = \sum_i \Tr T_0(a) b_i^t,
\end{equation}
where $b^t \equiv t(b)$ and $\sum_i a_i \otimes b_i$ is taken from the specified tensor cone.

As Conjecture \ref{2} claims the decomposability of the composition $T_1 \circ T_2$ the specified cone should be $C_d$.
To proceed with the study of Conjecture \ref{2} we will look more closely at the dual relations between the distinguished tensor cones. 
\begin{prop}
$C^d_d = C_{PPT}.$
\end{prop}
\begin{proof}
We have divided the proof into a sequence of observations:
\begin{enumerate}
\item $A \in C_d$ if and only if $A \in \{ \sum_{ij} a^*_i a_j \otimes b^*_ib_j \} \cap \{ \sum_{ij} a^*_i a_j \otimes t(b^*_i b_j) \}$
\item $Tr a_n\sigma_m Tr b_m \varrho_n = Tr a_n \sigma_m Tr \varrho_n b_m = Tr a_n \otimes \varrho_n \cdot \sigma_m \otimes b_m.$
\item $Tr a \otimes b \cdot c \otimes t(d) = Tr( ac) \cdot  Tr (bt(d)) = Tr( ac) \cdot  Tr (d(t(b))  \\
= Tr( ac) \cdot  Tr( t(b) d) = Tr a \otimes t(b) \cdot c \otimes d.$
\item $Tr ab \geq 0$ for all $b \geq 0$ implies $a \geq 0$.
\item $\sum_{mn,kl} Tr a^*_m a_n \otimes \varrho^*_m \varrho_n \cdot b^*_k b_l \otimes \sigma ^*_k \sigma_l \geq 0$
or
$\sum_{mn,kl} Tr a^*_m a_n \otimes \varrho^*_m \varrho_n \cdot b^*_k b_l \otimes t(\sigma ^*_k \sigma_l) \geq 0$.
\end{enumerate}
Hence
$$\sum_{mn,kl} Tr a^*_m a_n \otimes \varrho^*_m \varrho_n \cdot b^*_k b_l \otimes \sigma ^*_k \sigma_l \geq 0$$
or
$$\sum_{mn,kl} Tr a^*_m a_n \otimes t(\varrho^*_m \varrho_n) \cdot b^*_k b_l \otimes \sigma ^*_k \sigma_l \geq 0$$
what ends the proof.
\end{proof}

The principal significance of the next proposition is that it provides a criterion to settle the Conjecture \ref{2}. We emphasize that the condition given in the proposition below is necessary for the correct definition of the functional $\widetilde{T_1 \circ T_2}$.

\begin{prop}
\label{3.4}
Let $\cT_2$  be such that $\cT_2 C_d \subseteq C_p$ where $\cT_2 \equiv T_2 \otimes id$. Then, the Conjecture \ref{2} holds.
\end{prop}
\begin{proof}
Under the assumption, the composition $\cT_1 \circ \cT_2$ is well  defined on $C_d$ for any positive map $T_1$. 
In particular:
\begin{equation}
\widetilde{T_1 \circ T_2}(\sum_{ij} a^*_i a_j \otimes b^*_i b_j) = \sum_{ij} Tr T_1 \circ T_2 (a^*_i a_j) t(b^*_ib_j)) = \widetilde{T_1|_{\cT_2(C_d)}}(\cT_2(
\sum_{ij}a^*_i a_j \otimes b^*_i b_j)).
\end{equation}
As $\widetilde{T_1}$ is positive defined thus $\widetilde{T_1 \circ T_2}$ too.
Consequently,  the associated functional $\widetilde{T_1 \circ T_2}$ is positive defined on the cone $C_d$. Thus, the selection rule implies that $T_1 \circ T_2$ is a decomposable map.
\end{proof}

Further we note  $<C_p, C_i> \geq0$ as the corresponding cones are dual each other. Hence $<\cT_2 C_d, C_i> \geq0$ 
(for $\cT_2(C_d) \subseteq C_p$) and $<C_d, \cT_2^d C_i> \geq0$. Thus
\begin{equation}
\label{*}
\cT_2 C_d \subseteq C_p, \quad \rm{ implies} \quad  \cT^d_2 C_i \subseteq C_{PPT}
\end{equation}

Consider $\cT_1 \circ \cT^d_2$ where now $T_1$ is a PPT map. It is obvious that $\cT_1 \circ \cT^d_2$ is well defined on the cone $C_i$.
Thus, 
$$\widetilde{T_1 \circ T^d_2}(\sum_{ij} a^*_ia_j \otimes b^*_ib_j) = \sum_{ij} Tr T_1(T^d_2(a^*_ia_j))t(b^*_ib_j)\geq 0,$$
where $\sum_{ij} T^d_2(a^*_ia_j) \otimes b^*_ib_j \in C_{PPT}$.
As $T_1$ is by assumption PPT map, $\widetilde{T_1 \circ T_2}$  is positive defined.
Hence,  the selection rule implies that $T_1 \circ T^d_2$
is a superpositivity map. And this shows that the examined Conjectures  \ref{1} and  \ref{2}  are dual each other in the above sense.

What is left is to examine  the condition  $\cT_2C_d \subseteq C_p$. This will be done in the next Section, for finite dimensional case.

\section{Finite dimensional case}

In this section we indicate how techniques described in the previous section may be used to examine the relations (\ref{*}). 
In addition, important PPT mapping classes will be described and illustrative examples will be given.

 As  finite dimensional models are prevailing in  Quantum Information, from now on, again,  we make the assumption that the considered Hilbert space $\cH$ is finite dimensional. Needless to say the finite dimension of $\cH$ will be assumed only for simplicity. It is possible to proceed the study of the condition $\cT_2 C_d \subseteq C_p$ in more general setting. It would require the more careful analysis of trace class operators, continuity of corresponding functionals and maps. But, we will not develop this point here. The advantage of finite dimensional approach lies in the fact that it sheds some new light on the structure of the set of positive maps, in particular, on the origin of non-decomposable maps.

We have seen  that the condition (\ref{*}) is crucial and we wish to examine in details this condition.
To this end we will look more closely at the image $\cT_2( C_d)$ of $C_d$ where $\cT_2$ is a PPT map.  
We begin with a general result on PPT maps.
\begin{lem}
\label{lemat1}
If $\varphi$ is a PPT map then $\varphi \otimes t$ is a positive map on $B(\cH) \otimes B(\cH)$.
\end{lem}
\begin{proof}
It is easily seen that
$$(\varphi \otimes id) \circ (id \otimes t) = \varphi \otimes t = \varphi \circ t \circ t \otimes t = (\varphi \circ t \otimes id) \circ t \otimes t.$$
As $t \otimes t$ is a positive map, $t \circ \varphi$ as well as $\varphi \circ t$ are CP maps (cf  \cite{KMP}) and the composition of two positive maps is a positive map the claim is proved.
\end{proof}
Then we check at once that for PPT map $\cT$ one has $\cT C_{cp} \subseteq C_d$.

To proceed with the analysis of the images of the above mentioned cones with respect to PPT maps
we remind that any $a \in B(\cH)$ can be uniquely represented as
\begin{equation} 
\label{repre}
a = \sum_{k,l = 1}^{dim \cH} a_{kl} e_{kl},
\end{equation}
where  the system $\{e_{kl} \}_{k,l =1}^{dim \cH} \in B(\cH)$ consists of  matrix units (so $e_{kl}^* = e_{lk}, \ e_{kl} e_{mn} = \delta_{lm} e_{kn}, \ \sum_i e_{ii} = \mathbf{1}) $. Furthermore, as $\dim \cH < \infty$, we have used the identification: $\mathcal {F}_T(\cH)  = B(\cH)$.
The advantage of using the basis $\{e_{kl} \}$ lies in the fact that the general form of the cone $C_p$ can be rewritten as
\begin{equation}
C_p = \{\sum_i a_i \otimes b_i\} = \{\sum_{i,kl,mn} a_{kl}^ib_{mn}^i  e_{kl} \otimes e_{mn}\},
\end{equation}
where $a_i \geq 0$, $b_i \geq 0$ and similar conditions for matrices $[a^i_{kl}] \geq 0$ , $[b^i_{mn}] \geq 0$.

We recall, for details see \cite{Ingram}, \cite{Dom}, \cite{Qui}, that for given a closed convex cone $C$ in a Hilbert space $\cH$ there is the function which assign to each point $x$ in $\cH$ the nearest point of $C$ to $x$. Such function is called the projection of $\cH$ onto $C$.
The advantage of using the projection $P_{C_p}$  onto a convex cone $C_p$ lies in the fact that the condition $\cT_2 C_d \subseteq C_p$ implies 
\begin{equation}
P_{C_p} \circ \cT_2|_{C_d} = \cT_2|_{C_d}.
\end{equation}

Note that the above condition selects those mappings for which the PPT hypothesis has a positive answer. This observation gains in interest if we realize that the projection $P_{C_p}$ has a description. Namely, we have noted that the cone $C_p$ was described in terms of a basis. The uniqueness of expansion with respect to a basis implies, in Qui-Wang terminology, \cite{Qui} that $\{e_{kl} \otimes e_{mn} \}$ has no $\textit{positive relations}$.
By Qui-Wang formula \cite{Qui}
\begin{equation}
P_{C_p}(x) = \sum_{S} a_{kl}b_{mn} e_{kl} \otimes e_{mn}, \quad x \in B(\cH) \otimes B(\cH),
\end{equation}
for a unique subset $S$ of $\{(kl \times mn)\}$. The subset $S$ is specified by
\begin{equation}
\sum_S a_{kl}b_{mn} (e_{kl} \otimes e_{mn}, e_{ij} \otimes e_{pr} \geq (x, e_{ij} \otimes e_{pr}),
\end{equation}
and
\begin{equation}
\sum_S a_{kl}b_{mn} (e_{kl} \otimes e_{mn}, e_{ij} \otimes e_{pr} = (x, e_{ij} \otimes e_{pr}), \forall_{ (ij \times pr) \in S}.
\end{equation}

However, as the above description does not seem to be very effective when applied to general PPT mappings, we will now move on to the description of 
certain selected classes of PPT mappings.  We will start with the class of positive maps having the following structure:
\begin{equation}
\label{CP}
\{ E \circ T \}
\end{equation}
and
\begin{equation}
\label{CP1}
\{ T\circ E \}
\end{equation}
where $T$ is a completely positive map while $E$ is a conditional expectation on an abelian subalgebra. To legitimate the above definition
we note that such conditional expectations exist, see \cite{Nak}, \cite{Tak}, \cite{Kad}. Moreover, such conditional expectations are completely positive, see \cite{Ster}. Consequently,  (\ref{CP}) as well as (\ref{CP1}) define a class of completely positive maps. Next, let us note that transposition $t$ is a positive mapping. Furthermore, it is well known that if the positive mapping $t: \cA \to \cB$ where $\cA$ and $\cB$ are $C^*$  algebras and $\cA$ is an abelian algebra, then $t$ is also a completely positive mapping, see Proposition 3.9 in \cite{Tak}. Thus
\begin{cor}
Maps defined in (\ref{CP}) as well as in  (\ref{CP1}) are PPT maps.
\end{cor}
Our next task is to describe in detail the structure of  mappings defined by (\ref{CP}). We note that similar arguments apply to maps defined by (\ref{CP1}). Clearly, any completely positive map has a canonical, Choi-Krauss, form. Thus we should study a general form of the map $E$. To this end we recall:
\begin{prop}
\cite{HJ}, Proposition 2.6;
\textit{Let $\cA$ be an abelian involutive subalgebra of $M_n(\Cn)$. Let $\cP = \{p_1,p_2,...,p_r\}$ be the set of minimal projections in $\cA$. Then}
$$\cA = \oplus_{j=1}^r \Cn p_j.$$
\end{prop}

and
\begin{thm}
\cite{AGK}, Theorem 2.6, and \cite{AL}.
Let $\cA$ be a unital $C^*$-algebra, let $\cB \subseteq \cA$ be a unital $C^*$-subalgebra, and let $\varphi : \cA \to \Cn$ be a faithful tracial state. If $\cB$ is finite dimensional, then there exists a unique $\varphi$-preserving conditional expectation $P^{\varphi}_{\cB} : \cA \to \cB$ onto $\cB$ such that given any basis $\beta$ of $\cB$ which is orthogonal with respect to $<\cdot, \cdot>_{\varphi}$, we have
\begin{equation}
P^{\varphi}_{\cB}(a) = \sum_{b \in \beta} \frac{<a,b>_{\varphi}}{<b,b>_{\varphi}} b,
\end{equation}
for all $a \in \cA$.
\end{thm}
We will apply the above results to: $M_n(\Cn) \equiv B({\Cn}^n)$ and an abelian subalgebra $\cA \subseteq B({\Cn}^n)$. We may assume that $\cA$ has the set of minimal, mutually orthogonal projections $\cP = \{p_1,p_2,...,p_r\}$ such that $\sum_i p_i = \jed$.
We fix the tracial state $\varphi(\cdot) = \frac{1}{n} Tr(\cdot)$. Then the conditional expectation $E^{\varphi}_{\cP} (\cdot)$ has the form
\begin{equation}
\label{conexp}
E^{\varphi}_{\cP} (a) = \sum_{i=1}^r \frac{<a,p_i>}{<p_i>} p_i
\end{equation}
where $<a,b> = \frac{1}{n} Tr ab^*$.

It is worth pointing out that it is easy to check by direct calculations that $E^{\varphi}_{\cP}$ defined by (\ref{conexp})  has all properties characterizing a conditional expectation.

Let us consider the action of $E^{\varphi}_{\cP}\otimes id$ on $C_d$.
\begin{equation}
E^{\varphi}_{\cP}\otimes id(\sum_{ij} a^*_ia_j \otimes b^*_ib_j) = \sum_{ij} \sum_k \frac{<a^*_ia_j,p_k>}{<p_k>} p_k\otimes b^*_ib_j
=\sum_k p_k \otimes \sum_{ij} \frac{<a^*_ia_j,p_k>}{<p_k>} b^*_ib_j.
\end{equation}
It is easy to see that the image of $E^{\varphi}_{\cP}\otimes id$ is in $C_p$. Thus Proposition \ref{3.4} implies:
\begin{cor}
The maps defined by (\ref{CP}) satisfy the necessary condition
 for a positive answer to the PPT conjecture.
\end{cor}

The important issue concerning the description of the class of maps discussed here is the question of how large this class is. In other words, whether it contains non-trivial maps, for example, non-superpositive maps. It is important to note that the problem of defining non-superpositive PPT maps is dual to the problem of finding non-trivial nondecomposable maps. Therefore, our analysis sheds some new light on the old problem of the construction of nontrivial nondecomposable maps. 
To answer the above question and to illustrate our approach, we will consider in detail the first non-trivial case, namely the three-dimensional one.

\begin{exmp}
Let $\cA_1$ and $\cA_2$ be two abelian subalgebras in $B({\Cn}^3)$, where $\cA_1$ is generated by $\{e_{11} \equiv p_1, e_{22} \equiv p_2, 
e_{33}\equiv p_3 \}$ ($\cA_2$ by
$\{e_{11} \equiv p^{\prime}_1, e_{22}+e_{33} \equiv p^{\prime}_2 \}$ respectively) where $e_{ij} = e_i \otimes \bar{e_j}$, $\{e_i\}_1^3$ a basis in $\Cn^3$. Let us first consider maps related to algebra $\cA_1$.
Clearly, for the Choi matrix of $E^{\varphi}_{\cA_1}$, see \cite{Choi}, one has
\begin{equation}
[E^{\varphi}_{\cA_1}(e_{ij})]_{i,j=1}^3 =  [\delta_{j1} \delta_{i1} p_1 + \delta_{j2} \delta_{i2} p_2 + \delta_{j3} \delta_{i} p_3 ]_{i,j=1}^3.  
\end{equation}
It is easy to check that this form of Choi matrix  leads to superpositive maps $E^{\varphi}_{\cA_1}(A) = \sum_{i=1}^3 p_iAp_i$.
Now let us move on to the second case, i.e.  to maps associate with the subalgebra $\cA_2$.
The corresponding Choi matrix has the form
\begin{equation}
[E^{\varphi}_{\cA_2}(e_{ij})]_{ij=1}^3 =  [\delta_{j1} \delta_{i1} p_1^{\prime}]_{ij=1}^3 + \frac{1}{2} [(e_j, p_2^{\prime} e_i) p_2^{\prime}]_{ij \in \{1, 2,3\}}.
\end{equation}
However,  we note
\begin{equation}
(Trap_2^{\prime}) \cdot p_2^{\prime} = e_2 \otimes \bar{e_2} a e_2 \otimes \bar{e_2} + e_2 \otimes \bar{e_3}a e_3 \otimes \bar{e_2} + e_3 \otimes \bar{e_2}ae_2 \otimes \bar{e_3} + e_3 \otimes \bar{e_3}ae_3 \otimes \bar{e_3}.
\end{equation}
So even though the definition of the considered mappings seemed promising it is clear that this class of mappings is included in the set of superpositive mappings.
\end{exmp}

To find non-trivial PPT mappings for the three-dimensional case,  we will generalize maps defined in Example \ref{3.1}.
\begin{exmp} (Continuation of Example \ref{3.1})
We define
\begin{equation}
T_1(x) = \sum_{i=1}^4 V_i x V_i^*,
\end{equation}
where $V_1$ and $V_2$ are such as in Example \ref{3.1}, and $V_3 = f \otimes \overline{e_3} + Jf \otimes \overline{e_1}$ and
$V_4= Jf \otimes \overline{e_3} + f \otimes \overline{e_1}$ where $f$ is an arbitrary vector, but $f$ is neither a real nor an imaginary vector (and different from $v$)
in $\Cn^3$.
Proceeding analogously to Example 3.1, an easy calculation shows that $T_1 = t \circ T_1$. Thus $T_1$ is a PPT mapping. As mentioned earlier, the Choi-Krauss representation of CP mapping is not unambiguously determined. However, for the fixed Choi-Jamiolkowski matrix $[T(e_{kl})]_{kl}$ there is some improvement. To describe this we need the results taken from \cite{Choi} and \cite{Paulsen}:
\begin{itemize}
\item The Choi rank of CP map $T: M_n \to M_n$ is defined to be
\begin{equation}
cr(T) = \min \{ L: T(a) = \sum_{i=1}^L W_iaW^*_i \}.
\end{equation}
\item Theorem (Choi) If $T: M_n  \to M_n$ is CP map, $cr(T) = r$, and 
\begin{equation}
T(a) = \sum_{k=1}^r V_k a V^*_k = \sum_{l=1}^m W_l aW^*_l,
\end{equation}
then there exist unique $\alpha_{ij}$ such that
$$W_i = \sum_{j=1}^r \alpha_{ij}V_j.$$
\end{itemize}
Let us first consider the two-dimensional case. The above given results would imply:
\begin{equation}
\alpha_{11}V_1 + \alpha_{12}V_2 = h \otimes \overline{g},
\end{equation}
for some vectors $h,g \in\Cn^2$. But then
\begin{equation}
\label{2D1}
\alpha_{11} v \otimes \overline{e_1} + \alpha_{11} J v \otimes \overline{e_2} + \alpha_{12} Jv\otimes \overline{e_1} +\alpha_{12} v \otimes \overline{e_2} = h \otimes \overline{g}
\end{equation}
Therefore, equation (\ref{2D1})  and its Hermitian conjugate show that $h$ and $g$ are determined by $v$ and $Jv$. Consequently, the map $T$ is superpositive which is in perfect agreement with the analysis given in Example \ref{3.1}.

As the next step we will perform the same analysis for the mapping $T_1$, so now for the three-dimensional case. The superpositivity of the mapping $T_1$ would imply
\begin{equation}
\alpha_{11} V_1 + \alpha_{12}V_2 + \alpha_{13}V_3 + \alpha_{14} V_4 =   g\otimes \overline{w},
\end{equation}
for some $g,w \in \Cn^3$. This gives:
\begin{equation}
\label{3D}
\alpha_{11} v \otimes \overline{e_1} + \alpha_{11} J v \otimes \overline{e_2} + \alpha_{12} Jv\otimes \overline{e_1} +\alpha_{12} v \otimes \overline{e_2} + \alpha_{13} f\otimes \overline{e_3} + \alpha_{13} Jf\otimes \overline{e_1} 
\end{equation}
$$ + \alpha_{14} Jf\otimes \overline{e_3} + \alpha_{14} f\otimes \overline{e_1} = g\otimes \overline{w} .$$
The equation (\ref{3D}) and its Hermitian conjugate establish the relations between the vectors $v$, $f$, $g$ and $w$. For example:
\begin{equation}
\alpha_{11} v + \alpha_{12} Jv + \alpha_{13}f +\alpha_{14}Jf = (e_1,w)g,
\end{equation}
\begin{equation}
\alpha_{11}Jv + \alpha_{12} v = (e_2,w)g,
\end{equation}
and
\begin{equation}
\alpha_{13}f + \alpha_{14}Jf = (e_3,w)g.
\end{equation}
But $f$ and $v$ are arbitrary vectors  (only, neither real nor imaginary) thus the above relations can be satisfied for specific $v$ and $f$ but not for arbitrary vectors.
\end{exmp}

So we got:
\begin{cor}
The prescription for the map $T_1$ as well as its natural generalizations in higher dimensions leads to PPT maps which are not superpositive.
\end{cor}

The above discussion allows for a study of the key condition  $\cT_2 C_d \subseteq C_p$ given in the Proposition \ref{3.4}.
We first compute $T_1 \otimes id (C_d)$.
\begin{equation}
\sum_{ij}(T_1\otimes id)(a^*_ia_j\otimes b^*_ib_j) =
\end{equation}
$$\sum_{ij}[(a^*_ia_je_1,e_1) v\otimes \overline{v} + (a^*_ia_je_2,e_1) v\otimes \overline{Jv} +(a^*_ia_j e_1,e_2) Jv\otimes \overline{v} +(a^*_ia_j e_2,e_2) Jv\otimes \overline{Jv}] \otimes b^*_ib_j$$
$$+ \sum_{ij}[(a^*_ia_je_1,e_1) Jv\otimes \overline{Jv} + (a^*_ia_je_2,e_1) Jv\otimes \overline{v} +(a^*_ia_j e_1,e_2) v\otimes \overline{Jv} +(a^*_ia_j e_2,e_2) v\otimes \overline{v}] \otimes b^*_ib_j$$
$$+\sum_{ij}[(a^*_ia_je_3,e_3) f\otimes \overline{f} + (a^*_ia_je_1,e_3) f\otimes \overline{Jf} +(a^*_ia_j e_3,e_1) Jf\otimes \overline{f} +(a^*_ia_j e_1,e_1) Jf\otimes \overline{Jf}] \otimes b^*_ib_j$$
$$+\sum_{ij}[(a^*_ia_je_3,e_3) Jf\otimes \overline{Jf} + (a^*_ia_je_1,e_3) Jf\otimes \overline{f} +(a^*_ia_j e_3,e_1) f\otimes \overline{Jf} +(a^*_ia_j e_2,e_2) f\otimes \overline{f}] \otimes b^*_ib_j.$$

The above equation shows that for a large class of PPT mappings, not necessarily superpositive, it is difficult to determine whether the condition 
$\cT_2 C_d \subseteq C_p$ is satisfied. The main obstacle are terms like $(a^*_ia_je_2,e_1) v\otimes \overline{Jv}$. A workaround that allows us to specify mappings that satisfy the condition $\cT_2 C_d \subseteq C_p$ is the following modification of the formula for the $T_1$ mapping.
Define
\begin{equation}
T_0(a) = T_1(a) + T^{\prime}_1(a),
\end{equation}
where $T^{\prime}_1$ is define as $T^{\prime}_1 = \sum_{i=1}^4 W_i a W^*_i$ and
$W_1= v \otimes \overline{e_1} - Jv \otimes \overline{e_2}$,  $W_2= Jv \otimes \overline{e_1} - v \otimes \overline{e_2}$,
$W_3 = f \otimes \overline{e_3} - Jf \otimes \overline{e_1}$ and
$W_4= Jf \otimes \overline{e_3} - f \otimes \overline{e_1}$.
It is easily seen that for PPT map $T_0$, not necessary superpositive, the condition $\cT_2 C_d \subseteq C_p$ is satisfied.
Summarizing the given analysis, we can say that for certain PPT mappings, not necessary superpositive maps, the PPT conjecture does  hold. On the other hand, it seems doubtful that this conjecture holds for any PPT mappings.

\section{declaration}

\begin{itemize}
\item \textbf{Conflict of interest:} None.
\item \textbf{Funds:} No funds, grants, or other support was received.
\item \textbf{Data:} This manuscript has no associated data.

\end{itemize}


\end{document}